\documentclass[%
 reprint,
nofootinbib,
 aps,
floatfix,
]{revtex4-2}

\usepackage{amsfonts}
\usepackage{graphicx}
\usepackage{textcomp}
\usepackage{xcolor}
\usepackage{verbatim}
\usepackage{gensymb}
\usepackage{subfigure}
\usepackage{srcltx}
\usepackage{epsfig}
\usepackage{color}
\usepackage[tbtags,sumlimits,nointlimits,reqno]{amsmath}
\usepackage{amssymb}
\usepackage{float}
\usepackage[makeroom]{cancel}
\usepackage[normalem]{ulem}
\usepackage{url}
\usepackage{balance}
\usepackage{enumitem,kantlipsum}
\usepackage{graphicx}
\usepackage{dcolumn}
\usepackage{bm}
\usepackage{hyperref}

\newcommand{\E}{\mathbf{E}}

\newcommand{\B}{\mathbf{B}}

\renewcommand{\L}{\mathbf{\mathcal{L}}}

\newcommand{\epsr}{\epsilon^{\prime}}
\newcommand{\epsi}{\epsilon^{\prime\prime}}

\newcommand{\D}{\mathbf{D}}
\newcommand{\Hf}{\mathbf{H}}

\definecolor{amber}{rgb}{0.8, 0.33, 0.0}

\begin{document}

\preprint{APS/123-QED}

\title{Complex Eigenmodes and Eigenfrequencies in Electromagnetics}


\author{Jo\~{a}o G. Nizer Rahmeier}
\email{joaonizer@cmail.carleton.ca}
\author{Ville Tiukuvaara}%
\author{Shulabh Gupta}%
\homepage{http://www.doe.carleton.ca/~shulabh.gupta/}
\affiliation{%
 Department of Electronics, Carleton University \\ 
 Ottawa, Ontario, Canada
}%

\date{\today}

\begin{abstract}
A first comprehensive treatment on complex eigenmodes is presented for general lossy traveling-wave electromagnetic structures where the per unit length propagation phase shift ($\beta$) dependent complex eigenfrequencies $\Omega(\beta)$ are mapped to the frequency dependent complex propagation constant $\gamma(\omega_0)$ for variety of electromagnetic structures. Rigorous procedures are presented to first compute the complex eigenmodes of both uniform and periodic electromagnetic structures which are confirmed using full-wave simulations and known analytical results. Two mapping procedures are further presented for arbitrary uniform and periodic structures, where the known $\{\Omega-\beta\}$ relationship is expressed using polynomial and Fourier series expansions, respectively. Consequently replacing $\{\Omega,~j\beta\}$ with $\{\omega_0, \gamma\}$ in the known $\{\Omega-\beta\}$ relation, a characteristic equation is formed which is then numerically solved for the two unknowns, representing the physical dispersion relation $\omega_0(\beta)$ and the frequency dependent propagation loss $\alpha(\omega_0)$ of the structure. The mapping procedure is demonstrated for variety of cases including unbounded uniform media, rectangular waveguide, Drude dispersive metamaterial and a periodic dielectric stack, where exact propagation characteristics have been successfully retrieved in all cases across both passbands and stopbands across frequency. 
\end{abstract}

\maketitle


\section{Introduction}
\label{Sec: Introduction}
Electromagnetic metamaterials in the recent times have led to intense research on variety of artificial structures featuring exotic wave propagation characteristics with unprecedented control. They are constructed by periodically arranging sub-wavelength particles so that at the operating frequency, the unit cell period $p \ll \lambda$. In this regime, the structure can be homogenized and described using effective material parameters in terms of various constitutive parameters. By engineering these particles at this sub-wavelength scale, extensive macroscopic field control can be achieved to manipulate amplitude, phase and polarization of an electromagnetic wave in both space and time \cite{Caloz_MT_2009, Caloz_Wiley_2006, MTM_Eleftheriades}.

Typical metamaterials exhibit complex propagation regimes consisting of various frequency passbands and stopbands, in general. Metamaterials are thus closely related to the general class of periodic electromagnetic structures, where the unit cell period is not necessarily sub-wavelength \cite{Jackson_book_CED, Advanced_Engineering_EM, Rothwell_EM}. Fields propagating inside such structures within the passbands can be expressed in terms of space harmonics, and their respective contributions to the total fields strongly depend on the unit cell periodicity, $p$. The field propagation characteristics can be described using a frequency dependent complex propagation constant $\gamma(\omega_0) = \alpha(\omega_0) + j\beta(\omega_0)$, where $\beta(\omega_0)$ is the per unit length phase shift, $\alpha(\omega_0)$ is the per unit length attenuation experienced by the wave and $\omega_0$ is the frequency of the electromagnetic wave. They are typically obtained using a \emph{Driven-mode Analysis} where a finite sized electromagnetic structure is excited with a frequency $\omega_0$ and the corresponding phase shift and attenuation are measured \cite{Simualtor_Review,kashiwa1998fdtd}.

For \emph{lossless} periodic structures, while propagation constant $\gamma$ is purely imaginary and wave propagates without attenuation, it becomes purely real inside stopbands resulting in evanescent field along the structure. The frequency dependent complex propagation of an arbitrary periodic (with or without sub-wavelength periodicity) or a uniform structure can be obtained using \emph{Eigenmode Analysis}. It is performed on the unit cell element where a real phase shift $\beta p$ is applied across the unit cell along a specified direction. An eigenmode equation is formed using the Helmholtz wave equation, which is then solved for the corresponding eigenfrequencies and eigenmodes. For the case of a \emph{lossy} electromagnetic structure, $\gamma$ becomes complex inside both passbands and stopbands.  While the eigenfrequencies are purely real for lossless structures, they become complex for the lossy case, i.e. $\Omega(\beta) = \omega_r(\beta)+ j\omega_i(\beta)$ \cite{Gardiol_EM}. As opposed to driven mode analysis of a large finite-sized structure, the eigenmode simulation is computationally efficient, since it only requires a single unit cell of size $p$ while supposedly capturing the complete electromagnetic properties of the structure.

While the benefit of computing eigenmodes is quite evident, it is not clear how the complex eigenmodes $\Omega(\beta)$ can be used to extract the complex propagation constant $\gamma(\omega_0)$ of a general lossy structure. For the problem of lossless structures, $\{\alpha,~\omega_i\}=0$ so that $\omega_r(\beta) = \omega_0(\beta_0)$ is the desired dispersion relation of the structure, and represents a trivial case. Besides, the complex propagation constant $\gamma$ describes wave attenuation in space, while complex eigenfrequencies $\Omega$ describe losses as attenuation over time. They seemingly describe the propagation losses using very different physical phenomenon and one starts to wonder if these descriptions are even equivalent and can rigorously characterize the given electromagnetic structure. Beyond the fundamental nature of this question, there is also a strong practical significance where the problem of characterization of variety of traveling wave lossy electromagnetic devices \cite{pozar2011microwave, Rothwell_EM}, metamaterials, leaky-wave antennas \cite{Caloz_McrawHill_2011, LWA_Jackson,Otto_SB} and periodic structures in general, is critical in efficiently modeling systems of finite size.

Surprisingly, very little work has been done to explore the origins of complex eigenfrequencies in general lossy electromagnetic structures and media, and to investigate how they can be used to predict the corresponding driven responses. The problem can be more specifically stated as follows: Given the complex eigenfrequencies $\Omega(\beta)$ obtained using eigenmode analysis (analytically or using commercial electromagnetic simulation tools), how can we compute the corresponding complex propagation constant $\gamma(\omega_0)$, if this mapping exists? Some of the initial works were reported in early 1980's by Tsuji et al. in the analysis of dielectric resonators, where a specific mapping, $\omega_0 = |\Omega|$ and the Q-factor $Q = \omega_i/\omega_r$ has been used \cite{tsuji1982analytical, tsuji1983complex}\footnote{These relations are used in typical commercial electromagnetic simulators such as Ansys High Frequency System Simulator (HFSS) and CST Microwave Studio, for instance.}. However, this mapping is approximate and is only valid in passbands of general electromagnetic structures and applicable under low loss conditions. Some other works have recently been reported in the open literature which have tried to address this problem \cite{Dyab_ComplexO, Otto2012ComplexFV, King_EG_APS}, while questioning whether a unique mapping indeed exists between $\Omega$ and $\gamma$.

A unique mapping between complex frequencies $\Omega(\beta)$ and $\gamma(\omega_0)$ has finally been reported in \cite{Nizer_Gupta_EuCAP2020} for the canonical case of a rectangular waveguide, thereby indicating that it may possibly exist for an arbitrary electromagnetic structure. Shortly later, and very recently, a general mapping between complex spatial and temporal frequencies has been proposed applicable to arbitrary electromagnetic structures, beyond just canonical problems \cite{Caloz_Mapping}. The method rests on Taylor expansion of the known $\Omega(\beta)$, which is an analytic function and extends its domain using the method of analytical continuation to determine the complex propagation constant \cite{Kreyszig}.

However, the work in \cite{Caloz_Mapping} has several shortcomings. Firstly, the general mapping problem is not well-posed, where no distinction is made between the real part of the complex frequency, $\omega_r$ and the driving frequency $\omega_0$. This assumption that $\omega_0 = \omega_r$, greatly limits the validity of the proposed mapping, where for instance, the method fails to describe the stopband behavior of general electromagnetic structures including canonical cases of lossy uniform media and rectangular waveguides. Moreover, while applying the mapping procedure to a periodic structure problem, the contributions of the space-harmonics has been ignored and the periodicity of $\Omega(\beta)$ is not considered in the Taylor expansion procedure. The validity and limitations of the method has further not been identified and therefore, the method ceases to describe a rigorous relationship between $\Omega(\beta)$ and $\gamma(\omega_0)$ for a general lossy electromagnetic problem. As will be shown throughout this work, a distinction between $\omega_r$ and $\omega_0$ is at the core of the mapping between the two $\gamma-\Omega$ complex planes.

In this work, the concept of complex eigenfrequencies is finally demystified for arbitrary lossy electromagnetic structures including metamaterials and general periodic structures. The concept of complex eigenfrequencies is first explained using Maxwell's and the Helmholtz wave equations expressed in terms of $\Omega$. Subsequently, following the initial work of \cite{Nizer_Gupta_EuCAP2020}, the mapping between complex eigenfrequencies $\omega(\beta)$ and $\gamma(\omega_0)$ is gradually developed and extended to a general mapping procedure. In this method, the known eigenmode solutions, $\beta-\Omega$ relationship, is expressed using \emph{Taylor expansion} for the non-periodic cases, or  \emph{Fourier series expansion} for the periodic cases, to form a mathematical relationship between $\gamma$ and $\Omega$, which is then subsequently solved. A variety of cases are presented including canonical cases of uniform media, rectangular waveguide and a Drude dispersive metamaterial exhibiting frequency stopband where the proposed procedure is successfully applied. The concept of complex eigenfrequencies is further extended to general periodic structures using Floquet's theorem to establish the general form of the dispersion relation and Bloch impedance in terms of $\Omega$. Taking an example of an alternating dielectric stack structure, the proposed mapping procedure demonstrates the capability of the method to  rigorously obtain its passband and stopband characteristics.

The paper is structured as follows. Sec. \ref{Sec: 2 - Wave Prop.} presents the basic background of complex frequencies using Maxwell's equations, and describes the mapping problem between $\Omega(\beta)$ and $\gamma(\omega_0)$ in more details. Sec. \ref{Sec: 3 Uniform Structure} illustrates the $\Omega-\gamma$ relationship for two canonical problems \textemdash{} uniform media and rectangular waveguides \textemdash{} where closed-form expressions of the mapping are possible to obtain for better insight into this problem. It then extends the mapping to arbitrary uniform lossy electromagnetic structures and demonstrates the procedure using a Drude dispersive metamaterial with passband-stopband characteristics. The mapping procedure is then finally extended to general periodic structures using Floquet's theorem and illustrated using an example of an alternating dielectric stack in Sec. \ref{Sec: 4 Periodic Structure}. Finally, conclusions are provided in Sec. \ref{Sec: 5 Conclusion}.

\section{Complex Frequency (\texorpdfstring{$\pmb{\Omega}$}{Omega})\\ vs. \\ Complex Propagation (\texorpdfstring{$\pmb{\gamma}$}{gamma}) }
\label{Sec: 2 - Wave Prop.}

Consider the following time-domain Maxwell's equations describing the electromagnetic field propagation behavior in a generic \emph{lossy} medium:

\begin{align}
\nabla\times\mathbf{E} = - \frac{d\mathbf{B}}{dt},~\nabla\times\mathbf{H} =  \frac{d\mathbf{D}}{dt}
\end{align}

\noindent where $\{\D, \B\}$ are related to $\{\E,\Hf\}$ through the constitutive parameters. Assume that all field quantities $\E,~\Hf,~\D,~\B$ are expressed in terms of 

\begin{align}\label{Eq:FieldForm}
\psi(x,y,z,t) = \psi_0(x,y)e^{j\Omega t}e^{- \gamma z}
\end{align}

\noindent where $\Omega = \omega_r + j\omega_i$ and $\gamma = \alpha + j\beta$ are the complex frequency and complex propagation constants, respectively. Substituting this form in the above Maxwell's equations, we get

\begin{subequations}\label{Eq:ComplexMaxwell}
\begin{align}
&\nabla\times\mathbf{E}_s(\Omega) = - j\Omega \B_s(\Omega)\\
&\nabla\times\mathbf{H}_s(\Omega) =   j\Omega\D_s(\Omega)
\end{align}
\end{subequations}

\noindent where $\psi_s = \psi_0e^{-\gamma z}$, is the complex form of the fields, where now the constitutive form can be written in terms of complex frequency $\Omega$ as:

\begin{subequations}\label{Eq:ConstRela}
\begin{align}
\B_s(\Omega) = \mu(\Omega)\Hf_s(\Omega)\\
\D_s(\Omega) = \epsilon(\Omega)\E_s(\Omega).
\end{align}
\end{subequations}

\noindent Using \eqref{Eq:ConstRela} in \eqref{Eq:ComplexMaxwell} and eliminating $\Hf_s$ (or $\E_s$) leads to the Helmholtz wave equation in $\E_s$ (or $\Hf_s$) as

\begin{align}\label{Eq:HelmH}
\nabla^2\E_s - \Omega^2\epsilon(\Omega)\mu(\Omega)\E_s =0. 
\end{align}

\noindent This describes the evolution of the complex E-fields along the propagation direction $z$, in the complex $\Omega$ plane and admits field solutions of the form \eqref{Eq:FieldForm}. For example, if a TEM configuration for the fields is considered, we get the following complex propagation constant $\gamma$ and complex medium impedance $\eta_s$, satisfying the wave equation:

\begin{align}\label{Eq:ComGamma}
\gamma = j\Omega \sqrt{\epsilon(\Omega)\mu(\Omega)},~
\eta_s =  \sqrt{\frac{\mu(\Omega)}{\epsilon(\Omega)}} 
\end{align}

\noindent The propagation problem can now be classified into two categories:

\begin{enumerate}[leftmargin=*]
\item \emph{Driven Mode Analysis:} In this case a real driving frequency $\omega_0$ is used for the source, and complex propagation constant $\gamma$ of \eqref{Eq:ComGamma} becomes

\begin{align}
\alpha(\omega_0) + j \beta(\omega_0) = j\omega_0\sqrt{\epsilon(\omega_0)\mu(\omega_0)},
\end{align}

\noindent which describes the wave propagation in the complex $\gamma$ plane for a real driving frequency $\omega_0$. The instantaneous fields are of the form $\psi(x,y,z,t) = \psi_0(x,y)e^{-\alpha z}e^{j(\omega_0t - \beta z)}$ (considering forward propagating waves only, for instance). The amplitude of this wave decays ($\alpha>0$) along the propagation direction $z$, i.e. electromagnetic fields \emph{attenuate in space}, but remains constant with time at an arbitrary fixed position $z_0$. Driven mode analysis is thus used to compute the wave attenuation for a finite sized structure, which is excited with a source at one end.

\item \emph{Eigen-Mode Analysis:} In this case, a finite propagation length $p$ of a given medium is considered, and a constant phase shift, $(\beta\times p)$ is applied across it. Consequently, \eqref{Eq:ComGamma} becomes, 

\begin{align}\label{Eq:Complx}
\beta(\Omega) &= \Omega\sqrt{\epsilon(\Omega)\mu(\Omega)} 
\end{align}

\noindent which describes the wave propagation in the complex $\Omega$ plane for a real phase shift $\beta p$. The instantaneous fields are of the form $\psi(x,y,z,t) = \psi_0(x,y)e^{-\omega_i t}e^{j(\omega_r - \beta z) t}$. The amplitude of this wave decays ($\omega_i>0$) with time $t$, i.e. electromagnetic fields \emph{attenuate in time}, but remain constant throughout space at an arbitrary time $t_0$. Eigenmode analysis provides the wave propagation properties of an infinite structure and is useful in predicting the response of a large finite-sized structure using a smaller periodic section, of length $p$, of the structure\footnote{A uniform structure may be considered as a periodic structure with an arbitrary period.}.
\end{enumerate} 

\begin{figure}[!t]
	\centering
	\includegraphics[width=1\columnwidth,clip=true]{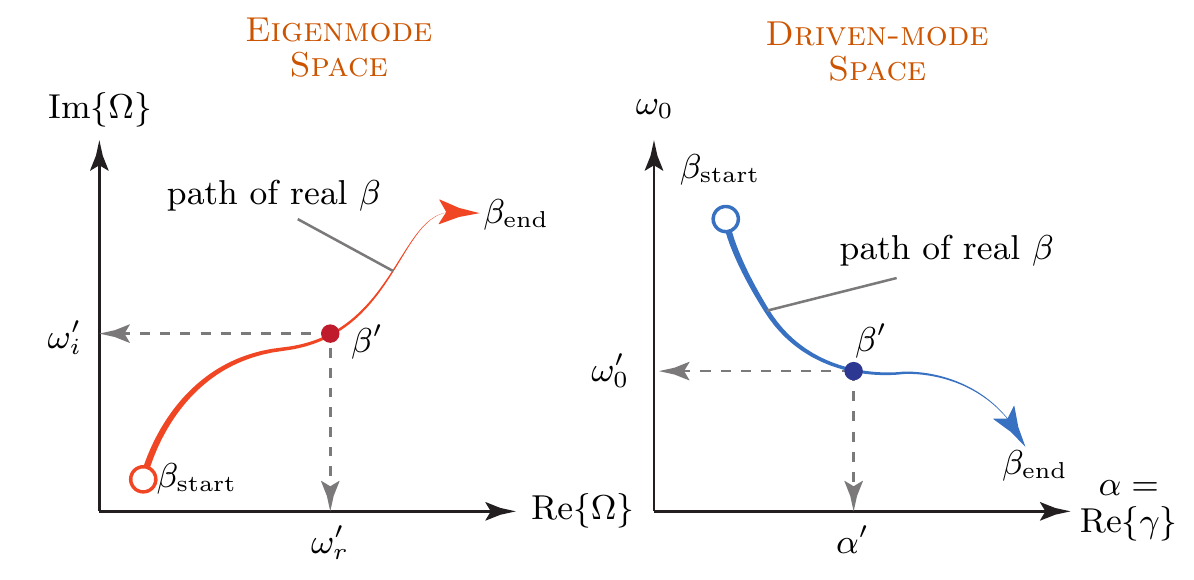}
	\caption{The problem of mapping a wave propagation described in a complex frequency $\Omega$ plane, to a modified complex propagation constant $\gamma$ plane.}
	\label{Fig:Planes}
\end{figure}

One can naturally ask: Are these two analysis techniques \textemdash{} driven and eigenmode \textemdash{} two independent approaches to solve electromagnetic wave propagation problems, or are they complementary where one analysis can be deduced from the other? For instance, can eigenmode analysis results of a small periodic unit cell be used to predict the driven mode response of a long finite-sized structure? From a design perspective, in eigenmode analysis, complex frequency $\Omega$ are computed for a specified real per unit length phase shift $\beta\in\{\beta_\text{start},~\beta_\text{end}\}$, i.e. $\Omega = f(\beta) = \omega_r + j\omega_i$. This corresponds to a known computed path in the complex $\Omega$ plane, tracing real values of $\beta$, as illustrated in Fig.~\ref{Fig:Planes}. The objective now is to find the unknown driving frequency $\omega_0$ of the wave which when excites the same structure, accumulates the same per unit length phase shift $\beta$, with the corresponding unknown per unit length attenuation constant $\alpha$, i.e. $\gamma = f(\omega_0) = \alpha + j\beta$. This can be better visualized using an alternate $\{\omega_0,\alpha\}$ complex plane as shown in Fig.~\ref{Fig:Planes} where the path traced in the $\Omega-$plane, maps to a path in this modified $\gamma-$space. Using several examples of various physical electromagnetic structures, we will show that this path is \emph{unique} for any periodic or uniform structure.

\section{Uniform Structures}\label{Sec: 3 Uniform Structure}

A more clear insight into the problem raised above can be obtained by considering the following two canonical cases of a bulk lossy medium and a rectangular waveguide, which will then be generalized to other structures in the following sections.

\subsection{Bulk Lossy Medium}

Consider an infinite unbounded lossy medium described by conductivity $\sigma_e$ and permittivity $\epsilon$ ($\mu = \mu_0$). The E-fields propagating in this medium at the driven frequency $\omega_0$ satisfying \eqref{Eq:HelmH} take the form

\begin{align}
\E = \E_se^{j\omega_0 t}e^{-\gamma z}
\end{align}

\noindent with the propagation constant $\gamma$ of \eqref{Eq:ComGamma} given by

\begin{align}\label{Eq:BulkGamma}
\gamma-\text{Plane}&:~\gamma^2 = (\alpha + j\beta)^2  = -\omega_0^2\mu \epsilon \left(1 - j \frac{\sigma_e}{\omega_0\epsilon}\right)
\end{align}

Alternatively, if \eqref{Eq:ComGamma} is restricted to be purely imaginary $\gamma$, i.e. ($\alpha=0$), the driven frequency $\omega_0$ must now become complex in order to satisfy \eqref{Eq:ComGamma} and the Helmholtz wave equations \eqref{Eq:HelmH}. This thus also represents a valid mathematical solution of the wave equation in the $\Omega$ space and corresponds to the eigenmode analysis of the medium. The E-fields are now given by

\begin{align}
\E = \E_se^{j\Omega t}e^{-j\beta z}
\end{align}

\noindent with the complex frequencies governed by 

\begin{align}\label{Eq:BulkOmega}
\Omega-\text{Plane}&:~\beta^2 = \Omega^2\mu \epsilon \left(1 - j \frac{\sigma_e}{\Omega\epsilon}\right)= f(\Omega)
\end{align}

\noindent In obtaining this expression, $\{\omega_0, \gamma\}$ have simply been replaced by $\{\Omega, j\beta\}$ in \eqref{Eq:ComGamma}, while keeping the material constants the same. It is important to note that $\omega_0\ne \text{Re}\{\Omega\}$, and they represent two independent variables. It is also instructive to consider the H-fields to complete the field solution. Applying the Maxwell-Faraday's law of \eqref{Eq:ComplexMaxwell}, we obtain $\Hf = (\nabla \times \E_s)/\eta$, where $\eta$ is the medium impedance given 

\begin{align}
 \eta &= \sqrt{\frac{\Omega\mu}{(\Omega\epsilon - j \sigma_e)}}
\end{align}

\noindent Next, separating real and imaginary parts of \eqref{Eq:BulkOmega}, we get the explicit expressions of the real and imaginary parts of the complex frequency $\Omega$ in terms of the real $\beta$ and material parameters as

\begin{align}\label{Eq:OmegaBulkCom}
\omega_r  =& \text{Re}\{\Omega\}= \pm\sqrt{\left(\frac{\beta^2}{\mu\epsilon}- \frac{\sigma_e^2}{4\epsilon^2}\right)},~ \omega_i = \text{Im}\{\Omega\}= \frac{\sigma_e}{2\epsilon}
\end{align}

\begin{figure*}[htbp]
	\centering
	\includegraphics[width=1\textwidth,clip=true]{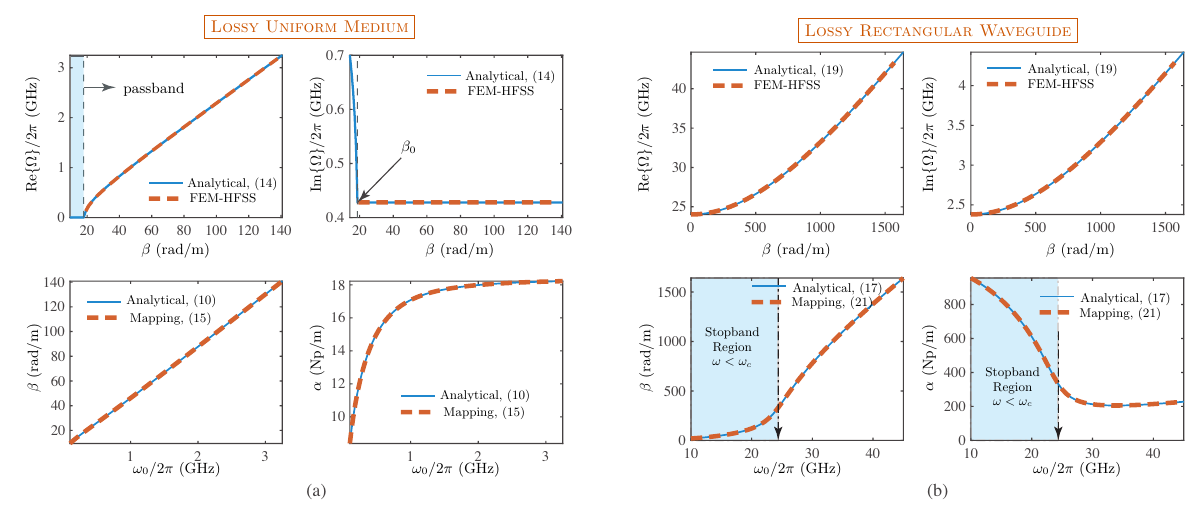}
	\caption{The complex eigenmodes $\Omega(\beta)$ and the mapped frequency dependent complex propagation constants for the case of (a) Lossy Bulk Medium with $\epsilon_r=4.2$ and $\sigma_e=0.2$~S/m, and (b) Lossy rectangular waveguides with $\epsilon_r=4.2$, $\delta=0.2$, and $a=3$~mm.}
	\label{Fig:BulkWG}
\end{figure*}

\noindent We note that $\omega_i$ is a constant with respect to $\beta$, while $\omega_r$ may cease to exist if the argument inside the square root becomes negative, and thus becomes imaginary. This imaginary contributions consequently adds to $\omega_i$. Therefore, a very lossy uniform media has a forbidden propagation band where $\beta < \beta_0$. Typical commercial simulators like Ansys HFSS (High Frequency System Simulator), can compute these complex eigenfrequencies. To validate \eqref{Eq:OmegaBulkCom}, an example of a very lossy medium is shown in Fig.~\ref{Fig:BulkWG}(a), and the numerically computed $\Omega(\beta)$ is compared with the above analytical expression. A perfect match is observed between the two. However, FEM-HFSS does not find any mode solution below $\beta_0$.

The next task is to predict the complex propagation constant $\gamma$ from this eigenmode response: more specifically, the dispersion relation $\omega_0(\beta)$ and the frequency dependent attenuation, $\alpha(\omega_0)$. Separating real and imaginary parts in \eqref{Eq:BulkGamma} allow us to find an equation for $\alpha$ and $\omega_0^2$. Applying the same complex expansion in \eqref{Eq:BulkOmega}, gives a description of $\beta^2$ as a function of $\Omega$. Combining all three equations allow us establish the desired mapping, so that 


\begin{subequations} \label{Eq:Bulk_Mapping}
\begin{align}
\omega_0(\beta) &= \pm2\beta\sqrt{\frac{\mu(\sigma_e\omega_i + \epsilon(\omega_r^2 - \omega_i^2))}{\mu^2\sigma_e^2+4\mu\epsilon\beta^2}}\\
\alpha(\beta) &= \frac{\mu\omega_0\sigma_e}{2\beta}
\end{align}
\end{subequations} 

\noindent where, different signs corresponds to various forward and backward propagating wave solutions. Using this mapping, the dispersion relation $\omega_0(\beta)$ and the frequency dependent attenuation, $\alpha(\omega_0)$ can now be readily computed from the eigenmode results (which in practice, are computed using numerical simulators). Fig.~\ref{Fig:BulkWG}(a) further shows these computed quantities and compares them with the known reference analytical result of \eqref{Eq:BulkGamma}. A perfect match is observed between the two, inspite of the fact that HFSS was not able to reproduce $\omega_i$ for $\beta<\beta_0$. This fully confirms the mapping procedure for this case of a uniform lossy media.

\subsection{Rectangular Waveguide}

Let us next consider the example of a lossy homogeneous rectangular waveguide of width $a$, filled with a dielectric material of permittivity $\epsilon=\epsr-j\epsi$ and permeability $\mu$. For the dominant TE$_{10}$ mode, the transverse electric field must satisfy the wave equation,

\begin{equation}\label{Eq:WaveRec}
    \frac{\partial^2E(x,y)}{\partial x^2}+k_c^2E(x,y)=0
\end{equation}

\noindent where $k_c=\pi/a$ \cite{sophocles2016orfanidis}. A 2D waveguide problem is assumed here for simplicity, so that field variation along the height dimension is considered zero. The analytical complex propagation constant, $\gamma$ of the dominant TE$_{10}$ mode is given by

\begin{align}\label{Eq:RecTE driven}
 \gamma-\text{Plane}:&~   (\alpha + j\beta)^2 = \omega_0^2\mu\epsilon - k_c^2
\end{align}

\noindent Following the same procedure as in bulk medium, we replace $\{\omega_0, \gamma\}$ with $\{\Omega, j\beta\}$, leading to

\begin{align}\label{Eq:RecEgnm}
\Omega-\text{Plane}:&~    \beta^2 = k_c^2 -\Omega^2\mu\epsilon = f(\Omega)
\end{align}

\noindent For specified per unit length propagation constant $\beta$, the complex frequencies may now be computed, which are found to be

\begin{subequations}\label{Eq: wr wi mapping}
    \begin{align}
        \omega_r&=\pm\frac{\delta\sqrt{2\varphi }}{2 \sqrt{\left(\delta ^2+1\right) \left(\sqrt{\delta ^2+1}-1\right)}}\label{eq: mat dep wr}\\
        \omega_i&=-\omega _r\frac{1\pm\sqrt{1+\delta ^2}}{\delta }\label{eq: mat dep wi}
    \end{align}
\end{subequations}

\noindent where $\varphi = (k_c^2+\beta^2)/\mu\epsr$, and $\delta=\epsi/\epsr$ is the loss tangent of the material inside the waveguide. The complex frequency $\Omega$  determined in \eqref{Eq: wr wi mapping} can be now verified with an eigenmode simulation of the waveguide. Fig.~\ref{Fig:BulkWG}(b) shows an example of a very lossy rectangular waveguide simulated in Ansys HFSS, whose eigenmode results are compared with the analytical expression of \eqref{Eq: wr wi mapping}. A perfect match is observed which validates the above analytical derivations.

Next, we isolate the material properties in \eqref{Eq:RecTE driven} and \eqref{Eq:RecEgnm}, and equate them as,

\begin{equation}
    \frac{k_c^2 - \gamma^2}{\omega_0^2} = \frac{k_c^2 + \beta^2}{\Omega^2}
\end{equation}
 
Now, separating real and imaginary parts and solving for the two unknowns $\omega_0$ and $\alpha$, allow us to establish a material-independent mapping\cite{Nizer_Gupta_EuCAP2020}, 

\begin{subequations}\label{Eq:RecMapping}
\begin{align}
\alpha(\beta) &= \frac{-\beta(\omega_r^2 - \omega_i^2) \pm \sqrt{\beta^2(\omega_r^2 + \omega_i^2)^2 +4\omega_r^2\omega_i^2k_c^2}}{2\omega_r\omega_i},\label{Eq: alpha Rec Independent Mapping}\\ 
\omega_0^2 &= \frac{(k_c^2 - \alpha^2 + \beta^2)(\omega_r^2 - \omega_i^2 ) + 4\alpha\beta\omega_r\omega_i}{(k_c^2 +\beta^2)}, \label{Eq: w0 Rec Independent Mapping}
\end{align}
\end{subequations}

\noindent Using the Ansys HFSS complex frequencies, and applying this mapping of \eqref{Eq:RecMapping}, the dispersion relation $\omega_0(\beta)$ and the frequency dependent attenuation, $\alpha(\omega_0)$ of a rectangular waveguide can now be obtained. These results are then compared with the known reference analytical result of \eqref{Eq:RecTE driven}, to fully confirm the mapping procedure. This comparison is also shown in Fig.~\ref{Fig:BulkWG}(b) which confirms that the mapping rigorously constructs the waveguide characteristics in both its passband and the stopband. It should be noted that while the eigenmode analysis finds modes above $\approx 23$~GHz only, the mapped results describes the waveguide behaviour accurately across all frequencies including $\omega_0\approx 0$.


\subsection{Mapping Procedure for Uniform Structures}

The methodology of computing complex frequencies $\Omega$, and their relationship with complex propagation $\gamma$ has so far been elucidated using canonical example of lossy bulk medium and rectangular waveguide. In each of the cases, $\beta^2 = f(\Omega)$ was known a priori, i.e. \eqref{Eq:BulkOmega} and \eqref{Eq:RecEgnm} for uniform lossy medium and rectangular waveguide, respectively. For a generic structure, these relations are not known in \emph{closed-form}, however $\Omega(\beta)$ computation is available using eigenmode simulations in typical numerical EM solvers. This information can thus be used to construct an analytical form describing the relationship between $\beta$ and $\Omega$.

Let us consider again the case of a uniform lossy medium and a lossy rectangular waveguide, described using complex frequencies:

\begin{subequations}\label{eq:genMed}
\begin{align}
\textit{Uniform Medium:}~\beta^2 &= -\Omega^2 \epsilon\mu = f_1(\Omega)\\
\textit{Rectangular Waveguide:}~\beta^2 &= k_c^2 -\Omega^2\epsilon  = f_2(\Omega)
\end{align}
\end{subequations}

\noindent In order to solve for the unknown parameters $\omega_0$ and $\alpha$, we replace $\{\Omega,~j\beta\}$ with $\{\omega_0,~\gamma\}$, and form the characteristic equation to be solved for the two unknowns. We note that for both these functions describing an unbounded and a bounded case, a square root function is involved, to describe the forward or backward propagating waves, separately. With this understanding, for any generic uniform (non-periodic or with sub-wavelength periodicity) medium for which $\{\beta, \Omega\}$ relationship is numerically know, we can construct a fitting function based on polynomial expansion, so that

\begin{align}\label{Eq:PolyExp}
\beta^2 = f(\Omega) = \sum_n^N a_n \Omega^n,~\Omega\in \mathcal{R},
\end{align}

\noindent where $\mathcal{R}$ is the region in the complex $\Omega$ plane where such an expansion is valid, and $f(\Omega)$ is the true unknown function for that specific medium. If the true function $f(\Omega)$ is analytic in the region $\mathcal{R}$, its polynomial expansion exists, and such an equivalence is valid \cite{Kreyszig}. Now following the previous methodology, we can simply replace $\{\Omega,~j\beta\}$ with $\{\omega_0,~\gamma\}$, so that

\begin{align}\label{Eq:Objec}
(\alpha + j\beta)^2 = \sum_n^N a_n \omega_0^n = f(\omega_0).
\end{align}

\noindent This is now the governing equation that establishes a mapping between the complex $\Omega$ space and the driven space quantities $\{\omega_0,~\alpha\}$, that now can be numerically solved, for each value of $\beta$. This general procedure may also be seen as an application of \emph{analytic continuation} procedure in complex plane analysis which is a well-known technique to extend the domain of a given analytic complex function \cite{Kreyszig}. It is clearly evident that $\beta^2(\Omega)$ of a lossy uniform medium and a bounded rectangular waveguide is a $2^\text{nd}$ order polynomial in $\Omega$, and thus equivalence of \eqref{Eq:PolyExp} is exact for all $\mathcal{R}$ in the complex $\Omega$ plane. We will next show how this procedure can operate on materials with more sophisticated constitutive frequency dispersion relations and corresponding propagation characteristics.

\subsection{Metamaterial with Drude Dispersion}

\begin{figure}[!htbp]
	\centering
	\includegraphics[width=\columnwidth,clip=true]{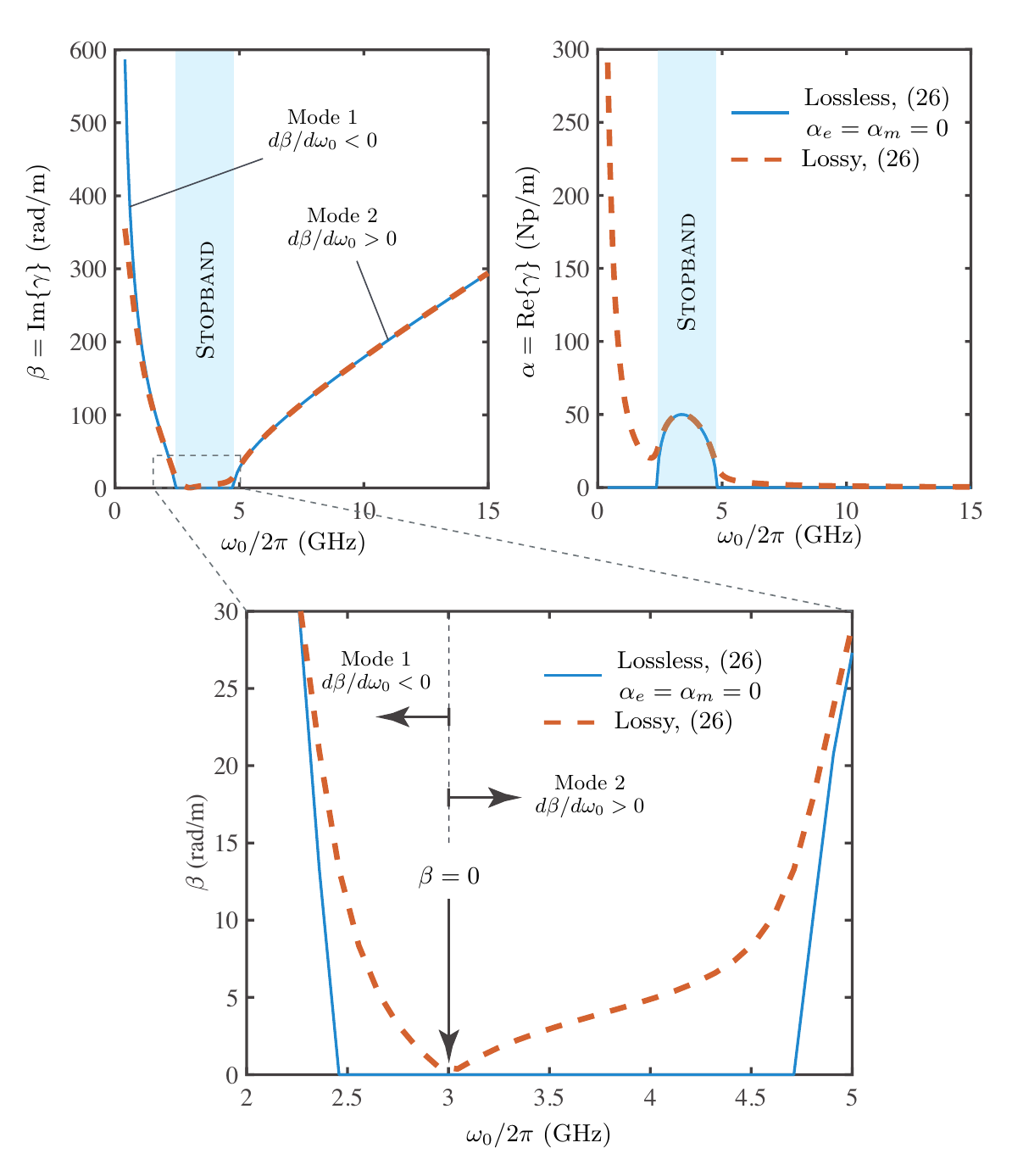}
	\caption{The analytical complex propagation constant $\gamma$ of a Drude dispersive metamaterial under lossy and lossless conditions. Design parameters are: $\omega_{ep}=15$~Grad/s, $\omega_{mp}=30$~Grad/s and $\alpha_e=\alpha_m=2\times10^9$.}
	\label{Fig:DrudeDriven}
\end{figure}

Let us consider an example of engineered metamaterials. Metamaterials are periodic artificial structures whose unit cell period $\Lambda$ is sub-wavelength, i.e. $\Lambda\ll\lambda_g$, where $\lambda_g$ is the guided wavelength of the electromagnetic wave \cite{Caloz_Wiley_2006}\cite{MTM_Eleftheriades}. Metamaterials are constructed using artificial scatterers which provide extensive dispersion control, where the effective constitutive parameters, $\mu(\omega)$ and $\epsilon(\omega)$ can be engineered. Consider for instance, an unbounded metamaterial, whose constitutive parameters within the frequency of interest are given by Drude dispersion

\begin{subequations}
\begin{align}
\epsilon(\omega) &= \epsilon_0\left(1 + \frac{\omega_{ep}^2}{j\omega\alpha_e - \omega^2}\right)\\
\mu(\omega) &= \mu_0\left(1 + \frac{\omega_{mp}^2}{j\omega\alpha_m - \omega^2}\right),
\end{align}
\end{subequations}

\noindent so that the complex propagation constant at a source frequency $\omega_0$ is given by 

\begin{align}\label{eq:gammaDrude}
\gamma^2 = \omega^2\epsilon(\omega_0)\mu(\omega_0)
\end{align}

 \noindent with $\{\omega_{ep},~\omega_{mp}\}$ as the electric and magnetic plasma frequencies and $\{\alpha_e,~\alpha_m\}$, the electric and magnetic loss coefficients, respectively. Fig.~\ref{Fig:DrudeDriven} shows a typical response of a Drude medium where the Drude parameters are chosen in such a way, that there is a large frequency stopband that occurs between the left-handed and right-handed propagation regions. For comparison, both lossless and lossy Drude medium characteristics are shown. The propagation constant $\gamma$ in the lossy case is complex in both passbands and the stopband, where the $\beta(\omega_0)$ branches corresponding to two left- and right-handed modes, respectively, converge to a single point somewhere inside the stopband at $\beta=0$. In contrast, the two modes are completely disconnected in the lossless case.
 
\begin{figure*}[!htbp]
	\centering
	\includegraphics[width=1\textwidth,clip=true]{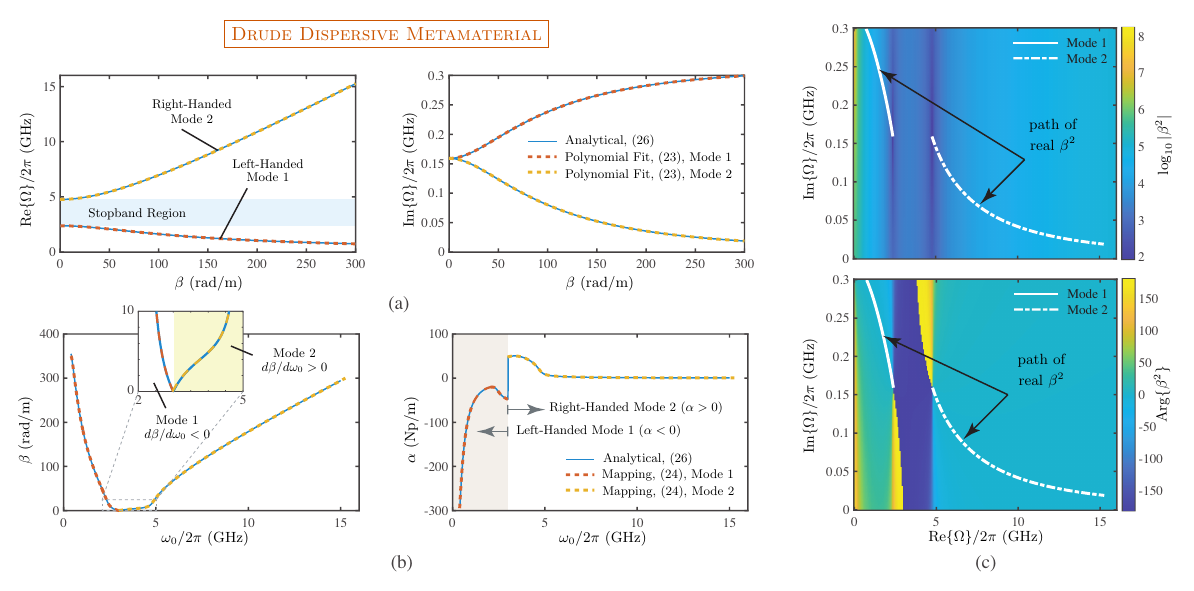}
	\caption{The complex eigenmodes $\Omega(\beta)$ and the mapped frequency dependent complex propagation constants for the case of a Drude dispersive metamaterial (a) complex frequencies $\Omega(\beta)$. (b) The driven mode dispersion relation $\beta(\omega_0)$, the frequency dependent attenuation constant $\alpha(\omega_0)$. (c) 2D map of $\Omega(\beta)$ showing the path of real $\beta$ computed using the eigenmode analysis. The design parameters are: $\omega_{ep}=15$~Grad/s, $\omega_{mp}=30$~Grad/s and $\alpha_e=\alpha_m=2\times10^9$.}
	\label{Fig:Drude_Mapping}
\end{figure*}

 To setup up the background of the mapping problem, we can first numerically compute its complex eigenfrequencies, by setting $\gamma = j\beta$ in \eqref{eq:gammaDrude}, and solve for $\Omega$. This step may also be performed in a numerical simulator such as Ansys HFSS, but the results will be identical. A typical complex frequency response is shown in Fig.~\ref{Fig:Drude_Mapping}(a) shown for the two modes.  The problem setup is complete, and $\beta-\Omega$ relationship is available. We now wish to extract its driven frequency response and compare to the analytical answer available in Fig.~\ref{Fig:DrudeDriven}.
 
The first step of the mapping procedure is to expand $\beta^2$ as a polynomial of $\Omega$ using the data available from the problem setup stage, following \eqref{Eq:PolyExp}. Since there are two modes (left-handed and right-handed), each mode is curve fitted separately using different polynomial expansions. This can be easily done in MATLAB using the function \texttt{polyfit()}, which returns complex coefficients $a_n$ for a specified order of the polynomial $N$. A good fit is observed in the frequency range of interest for both modes, as shown in Fig.~\ref{Fig:Drude_Mapping}(a), with degrees 5 and 7, respectively. Therefore, we now have $\beta^2(\Omega)$ relationship in terms of complex polynomial coefficients $a_n$. In the second and last step of the mapping procedure, the objective function of \eqref{Eq:Objec} is defined, and numerically solved for the two unknowns $\omega_0$ and $\alpha$, for each value of $\beta$. The computed results $\omega_0(\beta)$ and $\alpha(\beta)$ [or $\alpha(\omega_0)$] are shown in Fig.~\ref{Fig:Drude_Mapping}(b). To confirm if the mapping results are correctly obtained, a comparison is also shown in Fig.~\ref{Fig:Drude_Mapping}(b), with the known $\gamma(\omega_0)$ of a Drude medium of \eqref{eq:gammaDrude}. An excellent match is observed between the two, where both the passbands of left- and right-handed regions and the stopband characteristics are accurately reproduced. Moreover, to better capture the odd changes inside the stopband region we perform a piece-wise fit, with increased discrete resolution in this region. Both polynomial degrees were kept the same, 5 and 7 for left- and right-handed modes, respectively. 

To appreciate this example, we must realize that we are trying to approximate functions of the Drude form using a single polynomial of order $N$, i.e.

\begin{align}
\mu_0\epsilon_0\Omega^2\left(1 + \frac{\omega_{ep}^2}{j\Omega\alpha_e - \Omega^2}\right)\left(1 + \frac{\omega_{mp}^2}{j\Omega\alpha_m - \Omega^2}\right) \approx \sum_n ^N a_n\Omega^n
\end{align}

\noindent The left hand side of this equation is a ratio of two polynomials and exhibits two complex poles in the $\Omega$ plane for $\Omega=\{j\alpha_e,~j\alpha_m\}$. Fig.~\ref{Fig:Drude_Mapping}(c) shows the corresponding $\beta^2(\Omega)$ Drude function in the first positive quadrant of the complex frequency plane $\Omega$, where increasing values of $\beta^2$ are clearly observed along the imaginary axis, approaching a singularity. Therefore this function is not analytic around its pole region. Consequently, a polynomial fit is only valid in the complex plane region excluding these poles. Fortunately, these complex poles lie on the imaginary axis and are not within the region of interest. As a result, the polynomial expansion fit is sufficient to produce the correct driven frequency characteristics $\omega_0(\beta)$ and attenuation constant $\alpha(\beta)$. Fig.~\ref{Fig:Drude_Mapping}(c) further shows the path where $\beta^2$ (and thus $\beta$) is purely real. Both the left-handed and right-handed modes are clearly evident along with a forbidden stopband region, where no real $\beta$ exists.

This example of a metamaterial with Drude dispersion can now be easily extended to more sophisticated frequency dependent constitutive parameters or other bounded problems. For example, if a rectangular waveguide is filled with a Drude metamaterial, introduction of the cut-off wavenumber $k_c$, just alters the order of the polynomial of \eqref{eq:gammaDrude}, and the same mapping procedure can be successfully applied. We must remark that while a polynomial fit is a good approach to follow, it is not the only possibility to fit $\beta^2(\Omega)$. The complex frequencies of an arbitrary medium/structure may be expressed in terms of a mathematically equivalent Drude model in a specific region of $\Omega$, for instance. In that case, the characteristic equation $\beta^2 = f(\Omega)$ maybe solved in closed form.

\section{Periodic Structures}
\label{Sec: 4 Periodic Structure}

So far only uniform traveling-wave structures have been analyzed, which may be considered as periodic with arbitrary period. A more general class is that of periodic structures where the periodicity is unique and not necessarily sub-wavelength. In this section, we will extend the eigenmode analysis to periodic structures by describing the underlying methodology of computing complex eigenmodes using Floquet's theorem, and further extend the $\Omega-\gamma$ mapping procedure.

\begin{figure}[!t]
	\centering
		\includegraphics[width=0.9\columnwidth,clip=true]{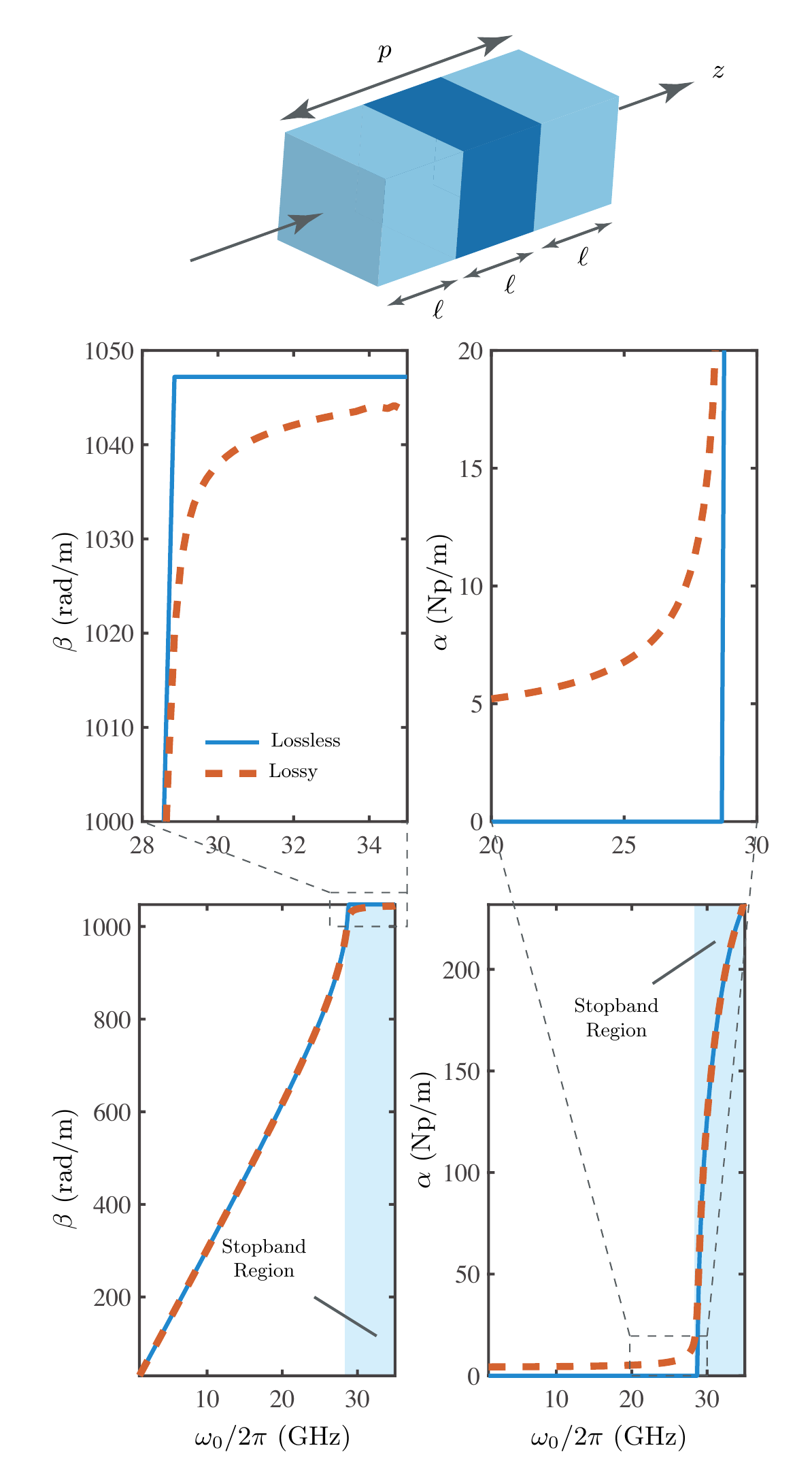}
		\caption{The analytical complex propagation constant $\gamma$ of a periodic dielectric stack under lossy and lossless conditions. Design parameters are: $\ell=1$~mm; Outer slabs: air; Inner slab: $\epsilon_r=4.2$ and $\sigma_e=0$~S/m for lossless and $\sigma_e=0.1$~S/m for lossy case.}
	\label{Fig:Periodic_Loss_Lossless}
\end{figure}

\subsection{Analytical Dispersion Relation using Floquet's Theorem}

Let us first consider a 1D uniform structure supporting a TEM mode for simplicity, where a forward propagating electric field may be expressed using $\gamma$ and $\Omega$ as
\begin{equation}
    \mathbf{E}(z,t)=E_0e^{-\gamma z}e^{j\Omega t}~\mathbf{\hat{x}} = E_x(z)e^{j\Omega t}
\end{equation}

\noindent where $E_x(z)=E_0e^{-\gamma z}$ is the complex form of the fields analogous to \eqref{Eq:ComplexMaxwell}. Moreover, the magnetic field, $\mathbf{H}(z,t)$, can be derived from Maxwell's curl equation, as
 \begin{align}
     \mathbf{H}(z,t)&=\frac{-j\gamma}{\Omega\mu}E_x(z)e^{j\Omega t}~\mathbf{\hat{y}}
 \end{align}

%
%
%
 
\noindent The propagation of the E- and H-fields  along $z$ and across a distance $\ell$ can be conveniently expressed as 

 \begin{subequations}\label{Eq: Fields Periodic}
     \begin{align}
     E_x(z+\ell)&=E_x(z)e^{-\gamma \ell}\notag{}\\
     &=E_x(z)\{\cosh(\gamma \ell)-\sinh(\gamma \ell)\}\label{eq: ex periodic}\\
     H_y(z+\ell)&=\frac{-j\gamma}{\Omega\mu}E_x(z)e^{-\gamma \ell}\notag\\
     &=\frac{-j\gamma}{\Omega\mu}E_x(z)\{\cosh(\gamma \ell)-\sinh(\gamma \ell)\}\label{eq: hy periodic}
 \end{align}
 \end{subequations} 
 
\noindent which can further be expressed using a compact transmission matrix form as 

\begin{equation}\label{Eq: Segement Relation}
    \begin{bmatrix}
    E_x(z+\ell)\\
    H_y(z+\ell)
    \end{bmatrix}=\begin{bmatrix}
    T_{11} & T_{12}\\
    T_{21} & T_{22}
    \end{bmatrix}\begin{bmatrix}
    E_x(z)\\
    H_y(z)
    \end{bmatrix}
\end{equation}

\noindent where,

\begin{equation}\label{Eq: Segment Transformation Matrix}
    \begin{bmatrix}
    T_{11} & T_{12}\\
    T_{21} & T_{22}
    \end{bmatrix} = \begin{bmatrix}
    \cosh(\gamma \ell) & -j\frac{\Omega\mu}{\gamma}\sinh(\gamma \ell)\\
    -j\frac{\gamma}{\Omega\mu}\sinh(\gamma \ell) & \cosh(\gamma \ell)
    \end{bmatrix}
\end{equation}

\noindent It should be noted that the transformation matrix elements are now expressed in terms of complex frequency $\Omega$.

Next consider a case of a periodic structure, as shown in Fig. \ref{Fig:Periodic_Loss_Lossless}. It is composed of three segments, where the outer slabs have same material properties $\epsilon_1$, $\mu_1$ and length $\ell_1$, and the inner slab has properties $\epsilon_2$, $\mu_2$ and length $\ell_2$, so that the cell period is $p=2\ell_1+\ell_2$. Using the same approach as of \eqref{Eq: Segement Relation} we can obtain the field relations for each individual slab. Then, by cascading \eqref{Eq: Segment Transformation Matrix} of each segment, one can build an overall transmission matrix of the entire unit cell as,

\begin{align}\label{Eq: Cell Transformation Matrix}
    \begin{bmatrix}
    T_{11}^\text{cell} & T_{12}^\text{cell}\\
    T_{21}^\text{cell} & T_{22}^\text{cell}
    \end{bmatrix}=&\begin{bmatrix}
    \cosh(\gamma_1 \ell) & -j\frac{\Omega\mu_1}{\gamma_1}\sinh(\gamma_1 \ell)\\
    -j\frac{\gamma_1}{\Omega\mu_1}\sinh(\gamma_1 \ell) & \cosh(\gamma_1 \ell)
    \end{bmatrix}\notag\\
    &\begin{bmatrix}
    \cosh(\gamma_2 \ell) & -j\frac{\Omega\mu_2}{\gamma_2}\sinh(\gamma_2 \ell)\\
    -j\frac{\gamma_1}{\Omega\mu_2}\sinh(\gamma_2 \ell) & \cosh(\gamma_2 \ell)
    \end{bmatrix}\notag\\
    &\begin{bmatrix}
    \cosh(\gamma_1 \ell) & -j\frac{\Omega\mu_1}{\gamma_1}\sinh(\gamma_1 \ell)\\
    -j\frac{\gamma_1}{\Omega\mu_1}\sinh(\gamma_1 \ell) & \cosh(\gamma_1 \ell)
    \end{bmatrix}
\end{align}

\noindent where $\gamma_n$ is the propagation constant of the $n^\text{th}$ segment. The fields on the surfaces at $z$ and $z+p$ are thus related through \eqref{Eq: Cell Transformation Matrix}.

Since the structure is periodic, we can now apply the Floquet theorem thereby relating the fields across the unit cell through a complex phase shift $\gamma$, so that $\psi(z+p) = \psi(z)e^{-\gamma z}$. Therefore, substituting \eqref{Eq: Cell Transformation Matrix} in \eqref{Eq: Segement Relation}, give us the following system of equations,

\begin{equation}
    \begin{bmatrix}
    T_{11}^\text{cell}-e^{-\gamma p} & T_{12}^\text{cell}\\
    T_{21}^\text{cell} & T_{22}^\text{cell}-e^{-\gamma p}
    \end{bmatrix}\begin{bmatrix}
    E_x(z)\\
    H_y(z)
    \end{bmatrix}=0
    \label{eq: ABCD generic dispersion}
\end{equation}
\noindent where for a non-trivial solution of the fields,
\begin{equation}
    \det{\begin{bmatrix}
    T_{11}^\text{cell}-e^{-\gamma p} & T_{12}^\text{cell}\\
    T_{21}^\text{cell} & T_{22}^\text{cell}-e^{-\gamma p}
    \end{bmatrix}} = 0.\notag
\end{equation}

\noindent For a reciprocal device, $T_{11}T_{22}- T_{12}T_{21}=1$, which leads to the following dispersion relation \cite{Otto_SB},

\begin{align}\label{Eq: Dispersion Relation Generic}
    \cosh(\gamma p) &= \cosh \left(2 \gamma_1\ell_1\right) \cosh \left( \gamma_2\ell_2\right) \notag\\
    &-\frac{(\eta_1^2+\eta_2^2)}{2 \eta _1\eta_2}\sinh \left(2\gamma_1\ell_1\right) \sinh \left( \gamma_2\ell_2\right)
\end{align}

\noindent where $\eta_n = (j\gamma_n/\mu\Omega)$ is the impedance of the $n^\text{th}$ segment. An important characteristics of a periodic structure is its characteristic impedance, also known as the \emph{Bloch impedance}, $Z_\text{B}$, which describes the input impedance of an infinite cascade of unit cells \cite{Otto_SB}. It can also be computed as\footnote{In Ansys HFSS eigenmode simulation, a custom field calculator function based on Maxwell's curl equation must be created to determine $H_y$ out of $E_x$ using the output variable \texttt{Freq}. A direct use of the $H_y$ field component does not consider the imaginary part of $\Omega$, therefore not satisfying Maxwell's equations.}

\begin{equation}\label{Eq: Bloch impedance}
   Z_\text{B}^{\pm} = \frac{\pm2T_{12}^\text{cell}}{(T_{22}^\text{cell}-T_{11}^\text{cell})\pm\sqrt{(T_{11}^\text{cell}+T_{22}^\text{cell})^2-4}} = \frac{E_x}{H_y}
\end{equation}

Let us proceed with this three layered dielectric structure and analyze \eqref{Eq: Dispersion Relation Generic} in both $\gamma$ and $\Omega$ planes. First, in the $\gamma$ plane, we select $\{\Omega,~\gamma\}$ to be $\{\omega_0,~\gamma\}$ in \eqref{Eq: Dispersion Relation Generic}, and based on the properties of each slab we are able to analytically compute $\alpha(\omega_0)$ and $\beta(\omega_0)$. For example, Fig. \ref{Fig:Periodic_Loss_Lossless}, shows the dispersion relation for a dielectric stack of air-dielectric-air, in a lossless condition ($\sigma_e=0$~S/m, inner slab) and a high loss condition ($\sigma_e=0.1$~S/m, inner slab). Due to its periodicity, the dispersion relation presents a stopband characterized by the abrupt jump of $\alpha$ and constant $\beta$ around 29 GHz. In a lossless condition, $\beta$ stays constant inside the stopband and $\alpha$ increases rapidly, attenuating the fields. Conversely, in a lossy condition $\alpha$ is finite in the passband, and, as depicted in the zoomed region of Fig. \ref{Fig:Periodic_Loss_Lossless}, there is a gradual transition of $\beta$ in the stopband.

\begin{figure}[htbp]
	\centering
	\includegraphics[width=1\columnwidth,clip=true]{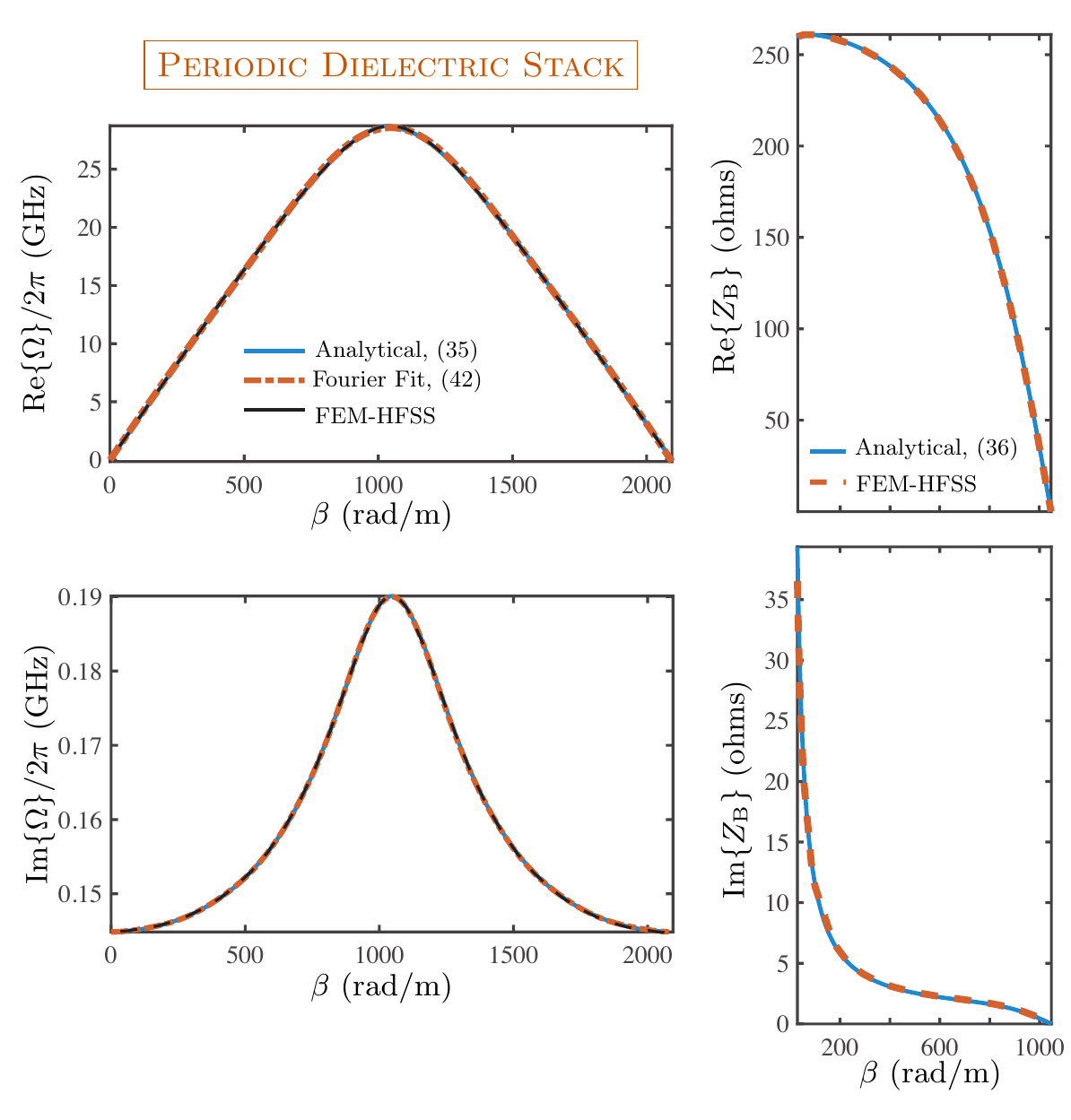}
	\caption{Complex frequency and Bloch impedance for for a lossy dielectric stack. Design parameters: $\ell=1$~mm; Outer slabs: air; Inner slab: $\epsilon_r=4.2$ and $\sigma_e=0.1$~S/m.}
	\label{Fig:Periodic_Omega_ZB}
\end{figure}

Another approach is to substitute $\{\Omega,~\gamma\}$ as $\{\Omega,~j\beta\}$, which corresponds to the $\Omega$ plane counterpart of \eqref{Eq: Dispersion Relation Generic}. This description gives the function $\beta(\Omega)$ that relates each complex frequency with a phase constant $\beta$. Using an optimization algorithm, one can determine the complex frequency, $\Omega$, by setting the objective function subject to $\text{Re}\{j\beta(\Omega)\}=0$ and $\text{Im}\{j\beta(\Omega)\}=\beta$. Such process is further used to compute the analytical eigenfrequencies of the dielectric stack, shown in Fig.~\ref{Fig:Periodic_Omega_ZB}. A complete agreement is obtained between the computed eigenfrequencies and the numerical solution of an Ansys HFSS eigenmode simulation. Besides, using the analytical values of $\Omega$ in \eqref{Eq: Bloch impedance}, also provides a perfect match between the analytical Bloch impedance and that extracted from the field information in HFSS eigenmode simulation, as also shown in Fig. \ref{Fig:Periodic_Omega_ZB}. The comparisons for $\Omega$ and $Z_B$ therefore validates the so far presented periodic equations for the dielectric stack and corresponding methodology.

\subsection{Periodic Mapping Procedure}

According to Floquet's theorem, the fields of a generic periodic structure of infinite length along the $z$-direction, can be represented using a periodic function,
\begin{align}\label{Eq:Floquet Theorem Series}
    \psi(x,y,z,\omega_0) & = \sum_{n=-\infty}^\infty\psi_n(x,y,\omega_0)e^{-\gamma_nz}
\end{align}
\noindent where,
\begin{equation}
    \gamma_n = \alpha +j\beta_n = \alpha +j\left(\beta+\frac{2n\pi}{p}\right)
\end{equation}
\noindent and,
\begin{equation}
    \psi_n(x,y,\omega_0)=\frac{1}{p}\int_{-p/2}^{p/2}\psi_p(x,y,\omega_0)e^{2\pi n z /p} dz.
\end{equation}

\noindent Each of the expansion terms inside the summation with $n\ne 0$ are known as the \emph{space-harmonics}. It should be recalled that they are not the modes of the structure, and only a complete sum of space harmonics representing the overall field is able to fulfill the Maxwell's equations.

A similar derivation can be obtained in the $\Omega$ plane for an eigenfrequency $\Omega$, except that in this case $\alpha=0$, and $\gamma_n=j\beta_n$, so that

\begin{align}\label{Eq:FloquetTheoremOmega}
\psi(x,y,z,\Omega) & = \psi_p(x,y,\Omega)e^{- j\beta z}
\end{align}

\noindent Consequently, the periodic description of the fields corresponds to a periodic relationship between $\Omega$ and $\beta$, such that, for a given mode, all space harmonics present the same $\Omega$ or the pair $\{\omega_r,\omega_i\}$, i.e.
\begin{align}
    \Omega(\beta) &= \Omega\left(\beta+\frac{2n\pi}{p}\right)
\end{align}

\begin{figure}[htbp]
	\centering
	\includegraphics[width=1\columnwidth,clip=true]{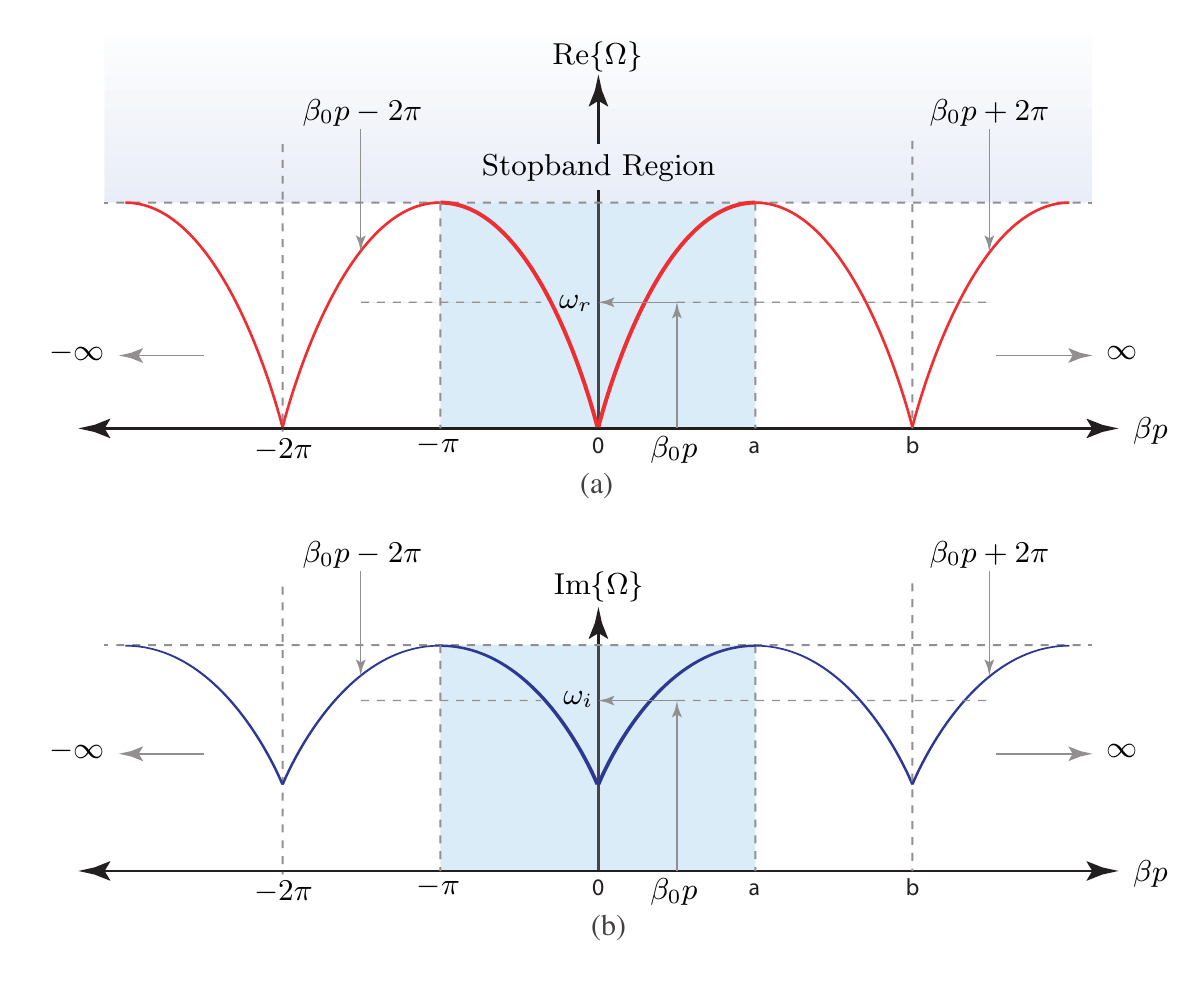}
	\caption{Band diagram of a general periodic lossy structure, illustrating (a) $\omega_r(\beta)$, and (b) $\omega_i(\beta)$. }
	\label{Fig:PeriodicBand}
\end{figure}

\noindent Therefore $\Omega$ is a periodic function of $\beta p $ with a period $2\pi$ for any generic periodic structure, as illustrated in Fig. \ref{Fig:PeriodicBand}. With this background, we now seek to obtain a general mapping procedure that can map $\Omega(\beta)$ of a generic periodic structure to $\gamma(\omega_0)$.

Consider a general periodic structure, of period $p$, for which $\Omega=f(\beta)$ relation is known or available using eigenmode simulations, where $\beta\in\{-\pi,\pi\}$ represents the Brillouin zone, i.e., the principal branch. While the analysis is done for $\beta p$ within $2\pi$ phase range (principal branch), the complete eigenmode solution consists of an infinite number of space harmonics, repeated at a period of $2\pi$. Consequently, the computed (or available) complex-frequencies are periodic functions of $\beta$. As a result, they can be expanded as a Fourier series:

\begin{align}\label{Eq: Omega Fourier Expansion}
    \Omega & = \omega_r(\beta)+j\omega_i(\beta) \notag\\
    & = \sum_{m=-\infty}^{\infty}A_m e^{jm(\beta p+2\pi)}+j\sum_{m=-\infty}^{\infty}B_n e^{jn(\beta p+2\pi)}
\end{align}

\noindent where $A_m$ and $B_n$ are complex coefficients used to fit $\omega_r$ and $\omega_i$, respectively. It is worth mentioning that, a polynomial Taylor expansion of $\Omega(\beta)$ would not preserve the periodic characteristic of the space harmonics for a thorough mapping between the $\gamma$ and $\Omega$ planes. Therefore, a complete description of $\Omega(\beta)$ for the entire $\beta$ range should rely on periodic basis functions, such as the Fourier series expansion. Once the Fourier coefficients are known, we can replace $\{\Omega,\beta\}$ in \eqref{Eq: Omega Fourier Expansion} by $\{\omega_0,-j\gamma\}$, so that

\begin{align}\label{Eq: omega  gamma periodic mapping}
    \omega_0 & = \sum_{m=-\infty}^{\infty}A_n e^{jm(-j\gamma p+2\pi)}+j\sum_{m=-\infty}^{\infty}B_n e^{jn(-j\gamma p+2\pi)}
\end{align}

\noindent Equation~\eqref{Eq: omega  gamma periodic mapping} can now finally be solved for the two unknowns, $\omega_0$ and $\alpha$ for each value of $\beta$ in the principal range. This process is characterized by a minimization problem, subject to $\text{Re}\{\omega_0(-j\gamma)\}=\omega_0$ and $\text{Im}\{\omega_0(-j\gamma)\}=0$. This completes the mapping procedure.

Let us now apply this procedure to the dielectric stack problem of Fig.~\ref{Fig:Periodic_Omega_ZB} and represent the known $\Omega(\beta)$ function using a Fourier series expansion. Before we begin, we note that for the first mode of this specific structure, shown in Fig.~\ref{Fig:Periodic_Omega_ZB} and Fig.~\ref{Fig:PeriodicBand}, $\beta=0$ gives $\omega_r=0$, which corresponds to a sharp transition between a forward wave ($v_p>0$), and a backward wave ($v_p<0$). More specifically, the $d\Omega/d\beta$ is discontinuous across $\beta=0$. This sharp transition may require an excessive increase in the fitting degree, which causes a serious impact in the mapping for lower frequencies of the first mode around this region, because of the Gibbs phenomenon. To overcome that, at every backward wave branch we curve fit the negative of the complex conjugate of $\Omega(\beta)$ instead, i.e. $-\Omega^*$ and exploit the symmetry of $\Omega(\beta)$ function about $\Omega=0$ axis. This mathematical transformation removes the sharp transition without inserting a discontinuity in $\omega_i(\beta)$, while preserving the field characteristics inside the periodic structure. Moreover, the negative of the complex conjugate only changes the time convention interpretation of the fields and are valid eigenmode solutions of the structure.

For the air-dielectric-air stack of Fig.~\ref{Fig:Periodic_Loss_Lossless}, Fig.~\ref{Fig:Periodic_Omega_ZB} also shows the resulting perfect fitting for $\Omega(\beta)$ using the Fourier series expansion in comparison with the analytical results and that provided by Ansys HFSS eigenmode simulation. For this case, we have used a Fourier series of degree 5 and 7 for $\omega_r(\beta)$ and $\omega_i(\beta)$, respectively.

\begin{figure*}[htbp]
	\centering
	\includegraphics[width=1\textwidth,clip=true]{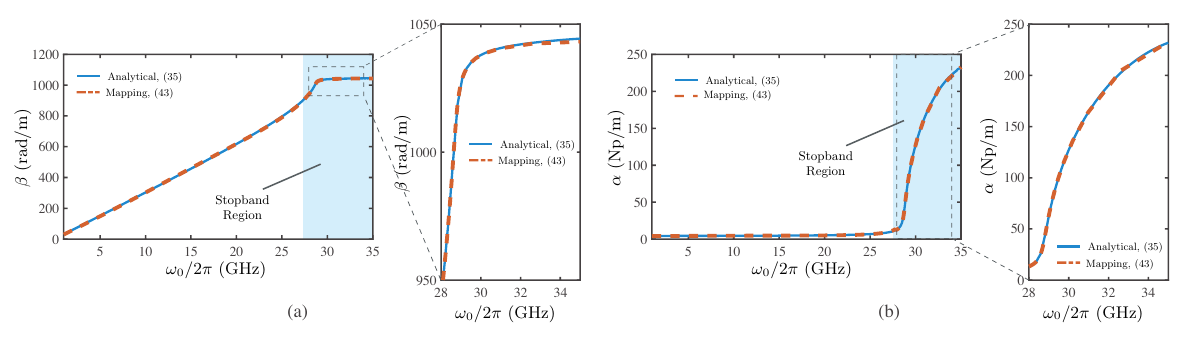}
	\caption{Extracted driven mode parameters using the proposed mapping for a periodic dielectric stack. (a) Dispersion relation $\beta(\omega_0)$. (b) Frequency dependent attenuation constant $\alpha(\omega_0)$. The design parameters are: $\ell=1$~mm; Outer slabs: air; Inner slab: $\epsilon_r=4.2$ and $\sigma_e=0.1$~S/m.}
	\label{Fig:Periodic_Mapping}
\end{figure*}

Next, we implement the optimization process using the \texttt{fminsearch()} available in Matlab, to determine $\alpha$ and $\omega_0$ in \eqref{Eq: omega  gamma periodic mapping} for each value of $\beta$. The mapped results are shown in Fig.~\ref{Fig:Periodic_Mapping}, where we are able to precisely reproduce the dispersion relation $\omega_0(\beta)$ and frequency dependent attenuation $\alpha(\omega_0)$ of the structure. The zoomed regions highlight the optimal matching between the $\gamma$ plane representation of \eqref{Eq: Dispersion Relation Generic} and the mapping result. While the mapping process is very robust in the passband region, the structures exhibits a large change in $\alpha$ and $\omega_0$ around the stopband with $\beta$ staying nearly constant (as can be also seen in Fig.~\ref{Fig:Periodic_Loss_Lossless}). Therefore, the mapping procedure becomes very sensitive to the fitting process and initial numerical guesses around the stopband region. Consequently, the results of Fig.~\ref{Fig:Periodic_Mapping} in the stopband region were obtained using piecewise fitting, where the stopband region was fitted using a different Fourier series expansion compared to that in the passband. The degrees were then reduce to 3 for both $\omega_r(\beta)$ and $\omega_i(\beta)$. This piece-wise fitting procedure nevertheless is still periodic and correctly takes the space contributions into account, thereby enabling the mapping to penetrate deeper into the stopband.

\section{Conclusions}
\label{Sec: 5 Conclusion}
A first comprehensive treatment of complex eigenmodes has been presented for general lossy traveling-wave electromagnetic structures where the complex eigenfrequencies $\Omega(\beta)$ are mapped to frequency dependent complex propagation constant $\gamma(\omega_0)$ for variety of electromagnetic structures. Rigorous procedures have been presented to first compute the complex eigenmodes of both uniform and periodic electromagnetic structures which have been confirmed using full-wave simulations. This established the methodology to compute and interpret the complex eigenfrequencies of generic lossy structures. Two mapping procedures have further been presented applicable to arbitrary uniform and periodic structures, where the known $\{\Omega-\beta\}$ relationship is expressed using polynomial and Fourier series expansions, respectively. In particular, the Fourier series expansion has been used to account for all the space-harmonic contributions in general periodic structures to obtain the correct mapping. Consequently replacing $\{\Omega,~j\beta\}$ with $\{\omega_0, \gamma\}$ in the known $\{\Omega-\beta\}$ relation, a characteristic equation is formed which is then numerical solved for the two unknowns, representing the physical dispersion relation $\omega_0(\beta)$ and the frequency dependent propagation loss $\alpha(\omega_0)$ of the structure. The mapping procedure has been demonstrated for variety of cases including unbounded uniform media, rectangular waveguide, Drude dispersive metamaterial and a periodic dielectric stack, where exact propagation characteristics have been successfully retrieved in all cases across both passbands and stopbands across frequency.

The proposed mapping procedure relies on mathematical fitting of the complex frequencies using polynomial and Fourier series expansions, which has been found to be particularly sensitive around the stopband regions, thereby requiring piecewise fitting process. Alternative methods based on closed-form descriptions of stopbands, e.g., an equivalent metamaterial, may alleviate this problem and future efforts maybe redirected in this direction. The presented methods and procedures to interpret complex eigenfrequencies, beyond its fundamental nature, has applications throughout the electromagnetic spectrum where it can describe the electromagnetic wave propagation in uniform and periodic structures such as antennas, optical waveguides and exotic metamaterials, for instance. Of particular importance is retrieving stopband characteristics of lossy radiating structures such as leaky-wave antennas, which is critical in understanding and solve the non-optimal broadside radiation problem in these structures \cite{Otto_SB}. While the analysis has been presented for traveling-wave type structures, it may also describe the complex resonant frequency of resonant structures and cavities, where the currently available expressions commonly used in typical commercial simulators are approximate and applicable under low-loss conditions only, while being limited to passband characteristics \cite{tsuji1982analytical,tsuji1983complex,Caloz_Mapping}. The current work thus lifts these limitations and presents a rigorous and an accurate mapping procedure to describe electromagnetic wave propagation in arbitrarily lossy structures and media.


\appendix

\bibliography{bib_complex_freq}

\end{document}